\documentclass[conference]{IEEEtran}
\IEEEoverridecommandlockouts
\usepackage{cite}
\usepackage{amsmath,amssymb,amsfonts}
\usepackage{algorithmic}
\usepackage{booktabs}
\usepackage{graphicx}
\usepackage{multirow}
\usepackage{textcomp}
\usepackage[table,xcdraw]{xcolor}
\usepackage{xspace}
\def\BibTeX{{\rm B\kern-.05em{\sc i\kern-.025em b}\kern-.08em
    T\kern-.1667em\lower.7ex\hbox{E}\kern-.125emX}}

\usepackage{url}
\usepackage{flushend}

\newcommand{\CC}{\cellcolor{gray!20}}
\newcommand{\tool}{\emph{DeepMatcher}\xspace}

\begin{document}
\title{Automatically Matching Bug Reports With Related App Reviews
\thanks{© 2021 IEEE.  Personal use of this material is permitted. Permission from IEEE must be obtained for all other uses, in any current or future media, including reprinting/republishing this material for advertising or promotional purposes, creating new collective works, for resale or redistribution to servers or lists, or reuse of any copyrighted component of this work in other works.}}


\author{\IEEEauthorblockN{Marlo Haering}
\IEEEauthorblockA{
\textit{University of Hamburg}\\
Hamburg, Germany \\
haering@informatik.uni-hamburg.de}
\and
\IEEEauthorblockN{Christoph Stanik}
\IEEEauthorblockA{
\textit{University of Hamburg}\\
Hamburg, Germany \\
stanik@informatik.uni-hamburg.de}
\and
\IEEEauthorblockN{Walid Maalej}
\IEEEauthorblockA{
\textit{University of Hamburg}\\
Hamburg, Germany \\
maalej@informatik.uni-hamburg.de}
}

\maketitle

\begin{abstract}
App stores allow users to give valuable feedback on apps, and developers to find this feedback and use it for the software evolution.
However, finding user feedback that matches existing bug reports in issue trackers is challenging as users and developers often use a different language.
In this work, we introduce \tool, an automatic approach using state-of-the-art deep learning methods to match problem reports in app reviews to bug reports in issue trackers.
We evaluated \tool with four open-source apps quantitatively and qualitatively.
On average, \tool achieved a hit ratio of 0.71 and a Mean Average Precision of 0.55.
For 91 problem reports, \tool did not find any matching bug report.
When manually analyzing these 91 problem reports and the issue trackers of the studied apps, we found that in 47 cases, users actually described a problem before developers discovered and documented it in the issue tracker.
We discuss our findings and different use cases for \tool.
\end{abstract}



\section{Introduction} \label{sec:introduction}
The app market is highly competitive and dynamic.
\emph{Google Play Store} and \emph{Apple App Store} offer together more than $\sim$4 million apps \cite{Statista:Online:2019} to users.
In this market, it is essential for app vendors to regularly release new versions to fix bugs and introduce new features \cite{McIlroy2016}, as unsatisfied users are likely to look for alternatives \cite{finkelstein2017investigating, Williams:App:2018}. 
User dissatisfaction can quickly lead to the fall of even popular apps~\cite{Li:Satisfaction:2010}. 
It is thus indispensable to continuously monitor and understand the changing user needs and habits for a successful app evolution.

\begin{figure}[t!]
    \centering
    \includegraphics[width=0.8\columnwidth]{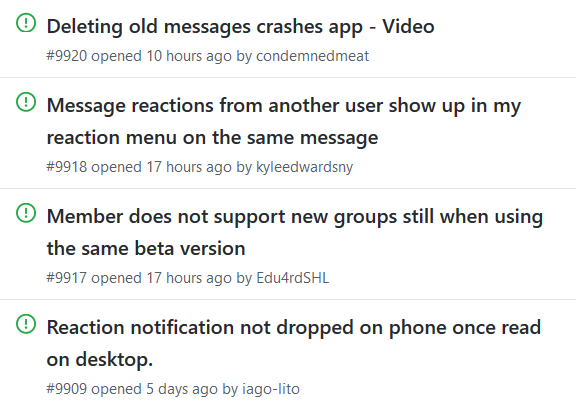}
    \caption{List of bug reports from the issue tracker of the app Signal Messenger.}
    \label{fig:bug_reports}
\end{figure}

However, identifying and understanding user needs and encountered problems is challenging as users and developers work in different environments and have different goals in mind.
On the one hand, software developers professionally report bugs in issue trackers to document and keep track of them, as illustrated in Figure \ref{fig:bug_reports}.
On the other hand, users voluntarily provide feedback on apps in, e.g., app reviews as shown in Figure \ref{fig:problem_report} --  using a different, often non-technical, and potentially imprecise language.
Consequently, seriously considering and using app reviews in software development and evolution processes can become time-consuming and error-prone. 

App vendors can regularly receive a large number of user feedback via various channels, including app stores or social media \cite{Pagano:App:2013, guzman2016needle}.
Manually filtering and processing such feedback is challenging.
In recent years, research developed approaches for filtering feedback, e.g., by automatically identifying relevant user feedback \cite{carreno2013analysis} like bug reports \cite{Stanik:Feedback:2019} and feature requests \cite{iacob2013retrieving}, or by clustering the feedback \cite{Villarroel:Release:2016} to understand how many users address similar topics \cite{Williams:Twitter:2017}.
While these approaches are helpful to cope with and aggregate large amounts of user feedback, the gap between what happens in the issue tracker and what happens online in the user space remains unfilled. 
For instance, developers remain unable to easily track whether an issue reported in an app review is already filed as a bug report in the issue tracker; or to quickly find a related bug they thought is already resolved.
Additionally, user feedback items often lack information that is relevant for developers, such as steps to reproduce or versions affected \cite{Zimmermann:TSE:2010, Martens:RE:2019}.

To address this gap, we introduce \tool, which is, to the best of our knowledge, the first approach that matches official and technically-written bug reports with informal, colloquially-written app reviews.
\tool first filters app reviews into \textbf{problem reports} using the classification approach by Stanik et al. \cite{Stanik:Feedback:2019}
Subsequently, our approach matches the problem reports with \textbf{bug reports} in issue trackers using deep learning techniques.
We use the state-of-the-art, context-sensitive text embedding method DistilBERT \cite{sanh2019distilbert} to transform the problem report and bug report texts into the same vector space. 
Given their vector embeddings, we then use cosine similarity as a distance metric to identify matches.

For 200 randomly sampled problem reports submitted by users of four Google apps, \tool identified 167 matching bug reports when configured to show three suggestions per problem report. 
In about 91 cases, \tool did not find any matches. 
We manually searched for these 91 cases in the issue trackers to check whether there are indeed no matching bug reports.
We found that in 47 cases, developers would have benefited from \tool, as no corresponding bug reports were filed.
We also qualitatively analyzed the context-sensitive text embeddings, which identified recurring bug reports and cases in which users reported problems before developers documented them.
We found that our approach can detect semantically similar texts such as ``draining vs. consuming battery'' and ``download vs. save PDF'', filling the gap between users' and developers' language.
Our qualitative analysis further revealed cases of recurring and duplicated bug reports. 
We share our replication package\footnote{\url{https://mast.informatik.uni-hamburg.de/replication-packages/}} for reproducibility.

The remainder of the paper is structured as follows.
First, we introduce \tool in Section \ref{sec:approach} explaining our design rationales. 
Section \ref{sec:study_design} introduces our evaluation setting, including the research questions, data, and process.
Section \ref{sec:results} presents our quantitative and qualitative evaluation results.
Then, we discuss how developers can use and modify \tool to detect bugs earlier and enrich existing issue descriptions with information extracted from user feedback in Section \ref{sec:discussion}.
Finally, we discuss the threats to validity in Section \ref{sec:validity}, 
related work in Section \ref{sec:related_work}, and conclude the paper in Section \ref{sec:conclusion}.

\section{Approach}\label{sec:approach}

\begin{figure*}[tb]
    \centering
    \includegraphics[width=\textwidth]{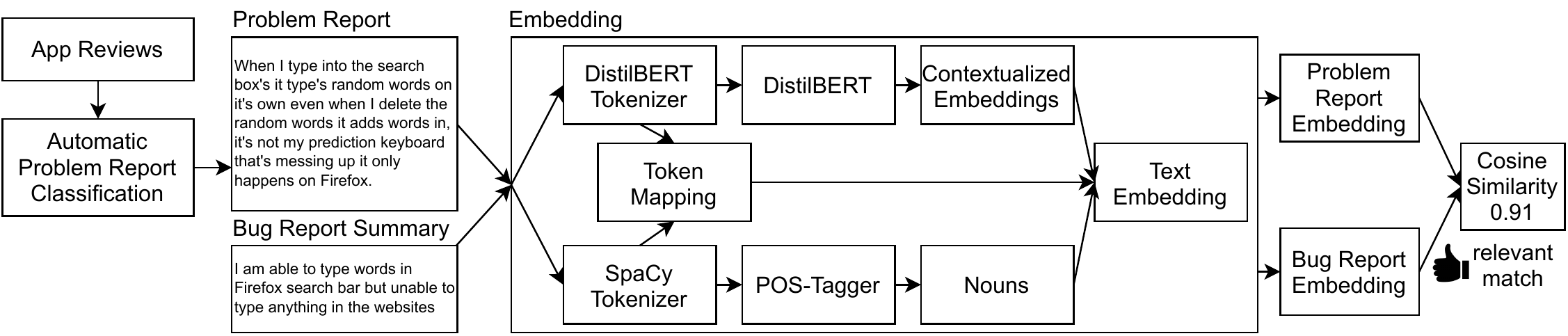}
    \caption{Overview of the \tool approach.}
    \label{fig:approach}
\end{figure*}

Figure \ref{fig:approach} shows an overview of \tool's technical approach.
The input of \tool is a problem report (an app review describing a problem with an app) and a bug report summary.
In Section \ref{ssec:filtering}, we discuss how we automatically identified problem reports from the review.
Section \ref{ssec:embedding} describes the text embedding creation process shown in the middle part of the figure.
This represents the transformation of textual data into numeric values, which we then use to calculate a similarity value as explained in Section \ref{ssec:matching}.

\subsection{Automatic Problem Reports Classification}\label{ssec:filtering}
\noindent\textbf{Challenges.}
One of the major problems when working with user feedback is the vast amount that software developers receive.
Particularly in app stores, Pagano and Maalej \cite{Pagano:App:2013} showed that developers of popular apps receive about 4,000 app reviews daily.
When considering Twitter as an alternative feedback source, Twitter accounts of popular software vendors receive about 31,000 tweets daily \cite{guzman2016needle}.
Besides the amount, the quality of the written feedback differs.
Most of the received app reviews simply praise, e.g., ``I like this app'' or dispraise, e.g. ``I hate this app!'' \cite{Pagano:App:2013}.
However, developers are particularly interested in the user experience, feature requests, and problem reports \cite{maalej2016automatic, Villarroel:Release:2016, guzman2017little}.
Our approach uses automatically classified problem reports from app reviews and subsequently matches them to bug reports in issue trackers.

\begin{figure}[t!]
    \centering
    \includegraphics[width=\columnwidth]{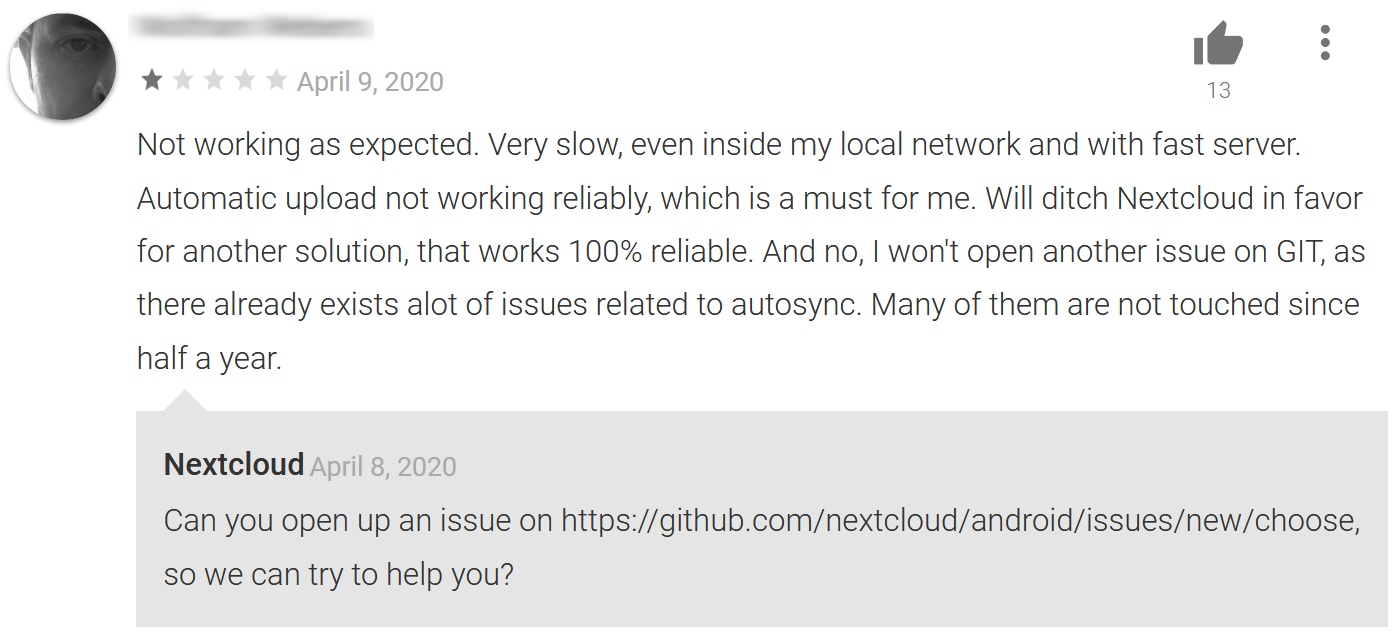}
    \caption{Example problem report for Nextcloud answered by the app developer.}
    \label{fig:problem_report}
\end{figure}

\noindent\textbf{Approach and Rationale.} 
We applied a four-step process to filter relevant app reviews.
First, we removed all user feedback containing less than ten words as previous research has shown that such feedback is most likely praise or spam \cite{Pagano:App:2013} and does not contain helpful information \cite{simmons2016agree}.
Second, we downloaded the replication package of Stanik et al. \cite{Stanik:Feedback:2019}, and applied the \emph{bug report}, \emph{feature request}, and \emph{irrelevant} classification approach to also filter the user feedback for bug reports.
The classification reduced the initial number of app reviews to 9,132 problem reports.
Fourth, to check the reliability of the classification, we randomly sampled and manually analyzed automatically classified app reviews for each of the four studied apps for manual analysis.
Two coders manually checked if the classified problem reports were correctly classified.
In case of disagreement, we did not include the app review but sampled a new one.
We repeated this step until we had 50 verified problem reports for each app, which is 200 in total.

\subsection{Text Representation with Word Embeddings} \label{ssec:embedding}
\noindent\textbf{Challenges.}
We further convert the text into a numerical presentation for further interpretation.
In natural language processing, practitioners usually transfer texts into vectors by applying techniques including bag-of-words, tf-idf \cite{manning1999foundations}, or fastText \cite{joulinBagTricksEfficient2017}.
When representing text in a vector space, we can perform calculations, such as comparing text similarity, which becomes essential in a later step for identifying matches.
Selecting the right word embedding technique is crucial, as it decides how precisely the vectors represent the text.
We face two major challenges.
First, users and developers usually use different vocabularies.
User feedback is more prone to spelling mistakes, often contain emoticons.
Moreover, users write mostly in an informal, colloquial, and non-technical way.
Second, bug reports are usually written in a more formal way, e.g., following templates, containing metadata, and may provide technical information like stack traces \cite {Zimmermann:TSE:2010}.

\noindent\textbf{Approach and Rationale.}
Both data sources consist of different text components.
While user feedback consists of a single text body, bug reports have a summary and a detailed description.
The description may contain a long explanation, including steps to reproduce, stack traces, and error logs.
We determined which text components of the bug report to include for the calculation of the text embedding.
Previous research showed that the detailed bug report description contains noise for information retrieval tasks \cite{zhou2016combining}. 
In particular, it contains technical details that users usually omit in their user feedback \cite{Martens:RE:2019}.
Further, research shows that the summary already contains the essential content of the long description \cite{ko2006linguistic, wang2008approach, lamkanfi2010predicting}.  
Therefore, we calculated the word embeddings only based on the bug report's summary.

Regarding the word embedding technique, we chose DistilBERT \cite{sanh2019distilbert}, a light-weight version of BERT \cite{devlinBERTPretrainingDeep2018} that is trained with a fewer number of parameters but has a similar generalization potential.
Alternative techniques would be, e.g., BERT, XLNet, or RoBERTa.
But as DistilBERT requires significantly fewer hardware resources and training time, it is more applicable for various development teams.
Our technique first tokenizes the input text and then calculates vectors for each token.
Compared to other text representations like bag-of-words or tf-idf, these vectors are contextualized; they consider the context of the surrounding words.
For example, the two sentences ``I love apples'' and ``I love Apple macbooks'' contain the token ``apple''. 
Contextualized embeddings take into account that the token's semantics differs in these two sentences.
In our approach, DistilBERT creates a 768-dimensional contextualized embedding for each token.

We calculated the document embedding from the individual token embeddings.
To reduce the weight of frequent but unimportant words such as ``I'', ``have'', or ``to'', previous research in text mining suggests removing stopwords \cite{panichella2015can, Villarroel:Release:2016, stanik2018simple}.
In our approach, we went one step further and only included embeddings of nouns, which we can automatically detect with a part-of-speech (POS) tagger.
We carefully decided to remove other parts of speech like the verb tokens as first trials showed that including frequent verbs like ``crash'', ``freeze'',  and ``hangs'' heavily biased our results toward these terms.
For example, \tool would match ``The app crashes when I open a new tab'' (problem report) with ``Firefox crashes on the home screen'' (bug report) because the verb ``crash'' puts the vectors of both texts closer together.
Based on this design decision, \tool weights essential words, i.e., nouns that describe components or features higher, while the contextualized token embeddings still contain information about the surrounding context, e.g., the verbs.
As a result, \tool, e.g., emphasizes the nouns ``new tab'' and ``home screen'' in the previous example and, therefore, would not consider the bug and problem report as a potential match.
Another positive side-effect of the surrounding context is that it helps to deal with misspelled words as their surrounding context is usually similar to the correct word's context.
Therefore DistilBERT calculates similar embeddings for them.

The automatic noun detection of the input texts is part of \tool and uses SpaCy's tokenizer and POS-tagger \cite{spacy}.
As SpaCy's tokenizer and the DistilBERT's tokenizer split the input text into different token sequences, we mapped the two sequences to each other by aligning them using pytokenizations.
For calculating the embedding for the full text of the problem report or bug report's summary, we added all noun word vectors of the text and averaged them.
Alternatively, we could have summed up the noun word vectors but decided to average them as the cosine similarity function depends on the vector angles and not on their lengths. 
Therefore, the choice of summing or averaging would not influence the cosine similarity score in our approach.

\subsection{Identifying relevant Bug Reports for a Problem Report} \label{ssec:matching}
\noindent\textbf{Challenges.} Given the numerical representation of the problem report and the bug report, \tool finally requires a method to decide whether a bug report is relevant for a problem report or not.
The main challenge in this task is calculating matching problem reports and bug reports with \emph{short text similarity} \cite{quan2010short}.
Besides semantic features, research tried text similarity approaches like simple substring comparisons \cite{islam2008semantic}, lexical overlap \cite{jijkoun2005recognizing}, or edit distances \cite{mihalcea2006corpus}.
Kenter and de Rijke \cite{kenter2015short} state that these approaches can work in some simple cases but are prone to mistakes.

\noindent\textbf{Approach and Rationale.}
We considered two options for this task.
One option is to model this task as a binary classification problem using the two classes, ``relevant'' or ``not relevant''.
However, this approach would require a large labeled dataset to train a classifier for this task, which is expensive and time-consuming \cite{cholletDeepLearningPython2018}.
Therefore, we chose the second option, which models this task as an information retrieval task.
Given a problem report as a query, we designed \tool to return a ranked list of suggested relevant bug reports.
We chose a distance function to measure the similarity between the two text embeddings and further rank the bug report summaries in decreasing order.

Two popular similarity measures for text embeddings are the euclidian similarity, and cosine similarity \cite{8947433, ghoshSimilarityBasedTextClustering2006}.
The euclidian distance can increase depending on the number of dimensions.
In contrast, the cosine similarity measures the angle of the two text vectors and is independent of their magnitude.
The benefit is that it results in a similarity value of 1 if the two vectors have a zero-degree angle.
A non-similarity occurs when the vectors have a 90-degree angle to each other.
Previous research \cite{phdthesis, ayataBUSEMSemEval2017Task2017, ayataTurkishTweetSentiment2017, ghoshSimilarityBasedTextClustering2006}, showed that cosine similarity performs equally or outperforms other similarity measures for dense text embedding vectors, which is why we also used it for \tool.
\section{Empirical Evaluation} \label{sec:study_design}
\subsection{Research Questions}
Apps usually receive user feedback as app reviews, which may contain the user's opinion and experience with the software.
Our study focuses on the app review category \emph{problem reports}, which is about users describing a faulty app behavior.
Figure \ref{fig:problem_report} shows an example of a problem report.
\emph{Bug reports} are issues in an issue tracker, complying with a certain template and contain information including summary, body, app version, timestamp, and steps to reproduce.
Figure~\ref{fig:bug_reports} shows a list of four bug report summaries of the Signal Messenger app in GitHub.

Our evaluation focuses on the following research questions:

\begin{description}
    \item[RQ1] \textbf{How accurate can \tool match app reviews with bug reports?} 
    \end{description}
    
In this research question, we analyze if we can identify bug reports in issue tracker systems for which users wrote app reviews. 
    For example, a developer filed the following bug report: \emph{``I am able to type words in Firefox search bar but unable to type anything in the websites''}.
    Can we find app reviews that describe the same issue, like the following app review? \emph{``When I type into the search box's it type's random words on it's own even when I delete the random words it adds words in, it's not my prediction keyboard that's messing up it only happens on Firefox.''}
  
  \begin{description}  
    \item[RQ2] \textbf{What can we learn from \tool's relevant and irrelevant matches?} 
\end{description}    

To answer this question, we checked a sample of relevant and irrelevant matches.
We analyzed cases in which contextual embeddings identify words with similar meanings, the language gap between developers and users, recurring bug reports, and a potential chronological dependency between problem reports and bug reports.
    We highlight our findings and explain them with examples from our dataset.

\subsection{Evaluation Data}
\label{sec:study_data}
For creating the evaluation data, we first collected app reviews and issues of four diverse apps.
We selected the apps Firefox (browser), VLC (media player), Signal (messenger), and NextCloud (cloud storage) to cover different app domains and usage scenarios in our analysis.
As Pagano and Maalej showed \cite{Pagano:App:2013}, most app reviews rather represent noise for software developers as they only contain praise like ``I like this app'' or insults like ``this app is trash''.
Therefore, we applied the bug report classification approach from Stanik et al. \cite{Stanik:Feedback:2019} to identify problem reports.
We chose this classification approach, as it uses state-of-the-art approaches, achieves high F1-scores of 79\% for English app reviews, and we could include the replication package in our pipeline without major modifications.
Eventually, we created a random sample from the collected data that we then used in the evaluation.

\begin{table}[]
\scriptsize
\caption{Overview of the Evaluation Data.}
\label{tab:study_data}
\begin{tabular}{@{}l|r|r|lrr@{}}
\toprule
App name         & \multicolumn{2}{c|}{Bug reports}                                                                        & \multicolumn{3}{c}{App reviews}                                                                                                                                                                                                                                            \\
                 & \multicolumn{1}{l|}{\begin{tabular}[c]{@{}l@{}}Time\\ period\end{tabular}} & \multicolumn{1}{l|}{Count} & \multicolumn{1}{l|}{\begin{tabular}[c]{@{}l@{}}Time\\ period\end{tabular}}                               & \multicolumn{1}{l|}{Count}                            & \multicolumn{1}{l}{\begin{tabular}[c]{@{}l@{}}Problem\\ reports \cite{Stanik:Feedback:2019}\end{tabular}} \\ \midrule
\rowcolor[HTML]{EFEFEF} 
Firefox Browser  & \begin{tabular}[c]{@{}r@{}}01/2011\\  08/2019\end{tabular}                & 29,941                      & \multicolumn{1}{l|}{\cellcolor[HTML]{EFEFEF}\begin{tabular}[c]{@{}l@{}}09/2018\\  07/2020\end{tabular}} & \multicolumn{1}{r|}{\cellcolor[HTML]{EFEFEF}5,706}  & 3,314                                                                                                     \\
VLC Media Player & \begin{tabular}[c]{@{}r@{}}05/2012\\  07/2020\end{tabular}                & 553                         & \multicolumn{1}{l|}{\begin{tabular}[c]{@{}l@{}}09/2018\\  07/2020\end{tabular}}                         & \multicolumn{1}{r|}{5,026}                          & 2,988                                                                                                     \\
\rowcolor[HTML]{EFEFEF} 
Signal Messenger & \begin{tabular}[c]{@{}r@{}}12/2011\\  08/2020\end{tabular}                & 7,768                       & \multicolumn{1}{l|}{\cellcolor[HTML]{EFEFEF}\begin{tabular}[c]{@{}l@{}}09/2018\\  08/2020\end{tabular}} & \multicolumn{1}{r|}{\cellcolor[HTML]{EFEFEF}10,000} & 2,583                                                                                                     \\
Nextcloud        & \begin{tabular}[c]{@{}r@{}}06/2016\\  08/2020\end{tabular}                & 2,462                       & \multicolumn{1}{l|}{\begin{tabular}[c]{@{}l@{}}06/2016\\  08/2020\end{tabular}}                         & \multicolumn{1}{r|}{774}                            & 247                                                                                                       \\ \midrule
\rowcolor[HTML]{EFEFEF} 
Total            &                                                                           & 40,724                      & \multicolumn{1}{l|}{\cellcolor[HTML]{EFEFEF}}                                                           & \multicolumn{1}{r|}{\cellcolor[HTML]{EFEFEF}21,506} & 9,132                                                                                                     \\ \bottomrule
\end{tabular}
\end{table}

As our study is concerned with matching problem reports found in app reviews with bug reports documents in issue trackers, we collected both---problem reports and bug reports.
In our study, we decided to evaluate our approach against four popular open-source Android apps, which stretch over different app categories.
We cover the categories browsing (Firefox), media player for audio, video, and streaming (VLC), a cloud storage client (Nextcloud), and a messaging app (Signal).
As these apps use different issue tracker systems to document bug reports, we developed crawlers for Bugzilla (Firefox), Trac (VLC), and GitHub (Nextcloud and Signal).
For each app, we collected all bug reports from the issue tracker systems.
As a requirement for our analysis, each bug report contains at least an ID, summary, and status (e.g., open and resolved), as well as the creation date.
Additionally, we also collected the remaining data fields provided by the issue tracker systems, such as issue descriptions and comments.
A complete list of the collected data fields is documented in our replication package.

We then collected up to 10,000 app reviews of the corresponding apps following Google's default sort order ``by helpfulness''.
Sorting by helpfulness helped us to not only considering the most recent app reviews (sort by date) but also emphasized the app reviews that other users deemed helpful.
For Nextcloud, we could not collect more than 774 app reviews as it seems that from their total 5,900 reviews in the Google Play Store, 5,126 reviews only contain a star rating without any text.
Our app review dataset covers a time frame of two to four years.
In total, we were able to collect 21,506 app reviews from the Google Play Store.
After applying the problem report classifier  \cite{Stanik:Feedback:2019}, we could reduce the number of app reviews to 9,132 problem reports.

Table \ref{tab:study_data} summarizes our study data.
The table reveals that while the time range of bug reports covers at least four years, Firefox has the highest number of bug reports filed from January 2011 to August 2019.
In total, we collected 40,724 bug reports, of which 29,941 belong to Firefox.
We focused on bug reports but ignored other issues like feature or enhancement requests by filtering the issues that developers labeled as such in the issue tracker systems.

\subsection{Evaluation Method} \label{sec:evaluation}
We evaluated \tool with respect to quantitative and qualitative aspects.
Starting from a set of manually verified problem reports, \tool suggested three bug reports for each.
We evaluated how accurately \tool finds matching bug reports based on their summaries for a given problem report.
We conducted a manual coding task, which consisted of two steps.

In the first step, we classified the app reviews using an existing approach \cite{Stanik:Feedback:2019} into problem reports to remove irrelevant feedback such as praise and dispraise (F1-score of 0.79 for English app reviews).
Then, we randomly sampled 50 problem reports per app and manually verified whether the classified app reviews are problem reports.
Two coders independently annotated the classification results according to a coding guide from previous research \cite{maalej2016automatic}.
We randomly sampled new problem reports until we reached 50 verified problem reports per app, which made 200 in total.

In the second step, we used \tool to calculate three suggestions of potentially matching bug reports for each of the 200 problem reports.
Again, two coders independently read each problem report and the three suggested bug reports.
For each matching, the coders annotated whether the match is relevant or irrelevant.
We consider the match relevant if the problem report and the bug report describe the same app feature (e.g., watch video) and behavior (e.g., crashes in full screen).
For example, for the problem report: \emph{``Latest update started consuming over 80\% battery. 
Had to uninstall to even charge the phone!''} \tool suggested the relevant bug report match \emph{``Only happening with latest version, But keep getting FFbeta draining battery too fast''}.
We documented the inter-coder agreement and resolved disagreements by having the two coders discussing each.
We report further analysis results based on the resolved annotations.

To answer RQ1, we calculated \tool's performance.
We report the number of relevant/irrelevant matches found per app and the mean average precision (MAP) \cite{zhu2004recall}.
It describes the average precision $AveP$ for each problem report $p$ and its suggestions and then calculates the mean over all problem reports $P$:

$$MAP =\frac{\sum_{p=1}^{P} AveP(p)}{P}$$

This is a conservative evaluation metric because it assumes that we have at least three relevant bug reports per problem report.
If this is not the case, even a perfect tool cannot achieve the highest average precision \cite{manning2009chapter}.
However, in our setting, the actual number of relevant bug reports is unknown.
Therefore, we additionally report on the hit ratio, which describes the share of problem reports for which \tool has suggested at least one relevant match.
For the irrelevant matches, we further tried to manually find relevant bug reports in issue trackers.
We further analyzed \tool's similarity score to identify a possible threshold, which users can use for the relevance assessment.

To answer RQ2, we conducted a qualitative analysis of the data.
For each app, we analyzed the language of app reviews and bug reports by counting the nouns used in both datasets in relation to the nouns used overall.
We highlight the strength of contextual word embeddings and show how \tool matches different words with similar semantic meaning.
We further analyze the cases in which developers report a bug report after a user submitted a related problem report in the app store.

\section{Evaluation Results} \label{sec:results}
\begin{table*}[]
\centering
\caption{Results of the Manual Coding for 4 Open Source Apps, each with 50 App Reviews. Legend: Mean Average Precision (MAP), number of suggested bug reports (\#).}
\label{tab:coding_agreement}
\begin{tabular}{l||r|r|r||r|r|r||rrrrr}
\hline
\multicolumn{1}{c||}{App} & \multicolumn{3}{c||}{1 Suggestion}                                                                                             & \multicolumn{3}{c||}{2 Suggestions}                                                                                            & \multicolumn{5}{c}{3 Suggestions}                                                                                                                                                                                                                                                                                                                                    \\
\multicolumn{1}{c||}{}    & \multicolumn{1}{c|}{\#} & \multicolumn{1}{c|}{MAP} & \multicolumn{1}{c||}{\begin{tabular}[c]{@{}c@{}}Hit\\ Ratio\end{tabular}} & \multicolumn{1}{c|}{\#} & \multicolumn{1}{c|}{MAP} & \multicolumn{1}{c||}{\begin{tabular}[c]{@{}c@{}}Hit\\ Ratio\end{tabular}} & \multicolumn{1}{c|}{\#}                          & \multicolumn{1}{c|}{MAP}                                        & \multicolumn{1}{c|}{\begin{tabular}[c]{@{}c@{}}Hit\\ Ratio\end{tabular}} & \multicolumn{1}{c|}{\begin{tabular}[c]{@{}c@{}}\#\\ Relevant\\ Matches\end{tabular}} & \multicolumn{1}{c}{\begin{tabular}[c]{@{}c@{}}Coder\\ Agreement\end{tabular}} \\ \hline
\rowcolor[HTML]{EFEFEF} 
Firefox                  & 50                      & 0.50                     & 0.50                                                                     & 100                     & 0.54                     & 0.58                                                                     & \multicolumn{1}{r|}{\cellcolor[HTML]{EFEFEF}150} & \multicolumn{1}{r|}{\cellcolor[HTML]{EFEFEF}0.58}               & \multicolumn{1}{r|}{\cellcolor[HTML]{EFEFEF}0.74}                        & \multicolumn{1}{r|}{\cellcolor[HTML]{EFEFEF}38}                                      & 0.93                                                                          \\
VLC                      & 50                      & 0.32                     & 0.32                                                                     & 100                     & 0.38                     & 0.44                                                                     & \multicolumn{1}{r|}{150}                         & \multicolumn{1}{r|}{0.40}                                       & \multicolumn{1}{r|}{0.51}                                                & \multicolumn{1}{r|}{26}                                                              & 0.91                                                                          \\
\rowcolor[HTML]{EFEFEF} 
Signal                   & 50                      & 0.38                     & 0.38                                                                     & 100                     & 0.47                     & 0.57                                                                     & \multicolumn{1}{r|}{\cellcolor[HTML]{EFEFEF}150} & \multicolumn{1}{r|}{\cellcolor[HTML]{EFEFEF}0.50}               & \multicolumn{1}{r|}{\cellcolor[HTML]{EFEFEF}0.68}                        & \multicolumn{1}{r|}{\cellcolor[HTML]{EFEFEF}45}                                      & 0.89                                                                          \\
Nextcloud                & 50                      & 0.62                     & 0.62                                                                     & 100                     & 0.73                     & 0.84                                                                     & \multicolumn{1}{r|}{150}                         & \multicolumn{1}{r|}{0.73}                                       & \multicolumn{1}{r|}{0.89}                                                & \multicolumn{1}{r|}{58}                                                              & 0.88                                                                          \\ \hline
\rowcolor[HTML]{EFEFEF} 
Total                    & 200                     & $\varnothing$ 0.45       & $\varnothing$ 0.46                                                       & 400                     & $\varnothing$ 0.53       & $\varnothing$ 0.61                                                       & \multicolumn{1}{r|}{\cellcolor[HTML]{EFEFEF}600} & \multicolumn{1}{r|}{\cellcolor[HTML]{EFEFEF}$\varnothing$ 0.55} & \multicolumn{1}{r|}{\cellcolor[HTML]{EFEFEF}$\varnothing$ 0.71}          & \multicolumn{1}{r|}{\cellcolor[HTML]{EFEFEF}167}                                     & $\varnothing$ 0.90                                                            \\ \hline
\end{tabular}
\end{table*}

This section reports the results of our evaluation study.
We analyze \tool's cosine similarity values to understand if we could use a certain similarity score threshold to identify matching problem reports and bug reports.
Further, we report on our qualitative analysis and describe relevant and irrelevant suggestions to find potential ways to improve automatic matching approaches.

\subsection{Matching Problem Reports with Bug Reports (RQ1)}\label{ssec:result_manual_analysis}
As introduced earlier, we sampled 50 problem reports per app (200 in total) and applied \tool to suggest matching bug reports.
In the first step, \tool suggested one matching bug report per problem report.
Then, we changed that parameter and let \tool suggest two matching bug reports.
Finally, \tool suggested three matching bug reports per problem report, which led to 600 suggestions.
Since \tool suggests bug reports based on the highest cosine similarity, it added one additional suggestion per step while keeping the previous ones.
This way, we could evaluate \tool's performance based on this parameter (number of suggestions).
Two authors independently annotated each of the 600 bug report suggestions as either a relevant or irrelevant match.

Table \ref{tab:coding_agreement} summarizes the overall result of the peer-coding evaluation.
The table shows that the inter-coder agreement for the whole dataset (3 suggested bug reports per problem report) is $\ge0.88$.
From the 600 matching bug report suggestions, the two coders identified 167 developer relevant matching suggestions.
These 167 suggestions occurred in 109 problem reports with the parameter \emph{number of suggestions} set to three.
Multiple relevant matches occurred either for generic problem reports like ``the app crashes often'' or for similar, recurring, or duplicated bug reports in the issue tracker.

\vspace{6pt}
\textbf{Suggestions Without Relevant Matches}.
For 91 problem reports, \tool could not find a relevant match within the three suggestions.
The reason for this is twofold: either no relevant bug report actually exists in the issue tracker system, or \tool missed relevant matches.
To understand why \tool did not identify any matches for 91 problem reports, we manually searched the issue tracker systems by building a query using different keyword combinations from the problem reports.
For example, Table \ref{tab:examples} shows a problem report of VLC for which \tool could not find a relevant matching bug report.
However, in our manual check, we found the bug report ``When the device's UI language is RTL, no controls are shown in the notification card'', which the two coders consider a relevant match.
For 47 problem reports, we could not find any relevant match in the issue tracker system, while \tool missed potentially relevant matches in 44 cases.
Consequently, \tool identified 47 problem reports that were undocumented in the issue trackers. This can help developers create new bug reports.

\vspace{6pt}
\textbf{Average Mean Precision and Hit Ratio}.
We calculated the Mean Average Precision (MAP) and the hit ratio of our manual annotated data for all three parameters (one suggestion, two suggestions, and three suggestions).
The MAP is a conservative score, which assumes that each problem report has at least as many relevant bug reports in the issue tracker as the parameter states.
For example, if we set the parameter for the number of suggested bug reports to three, the MAP score assumes that at least three relevant matching bug reports exist.
In case the problem report has less than three relevant bug reports, the average precision for that problem report cannot get the maximum value of one \cite{manning2009chapter}.
For our calculation, we excluded the problem reports for which we could not find a relevant bug report manually.
The hit ratio, on the other hand, is the number of problem reports for which \tool found at least one relevant match divided by the number of all problem reports.

Table \ref{tab:coding_agreement} shows the MAP and the hit ratio scores for each parameter setting.
Increasing the parameter from one to two shows that the MAP score increases by 8\%, while the hit ratio increases by 15\%, which means that we increase the chance of finding a relevant match to 61\%.
When further increasing the parameter to three, we observe that the probability of having at least one relevant match increases to 71\%, however as the MAP score reveals, developers might have to consider more irrelevant matches.
We found that for Nextcloud, \tool achieved the highest Mean Average Precision (0.73) and Hit Ratio (0.89).
In contrast, VLC achieved the lowest scores with a MAP of 0.40 and a hit ratio of 0.51.
Averaged over all apps, \tool achieved a mean average precision of 0.55 and a hit ratio of 0.71.

\begin{figure}[tb]
    \centering
    \includegraphics[width=\columnwidth]{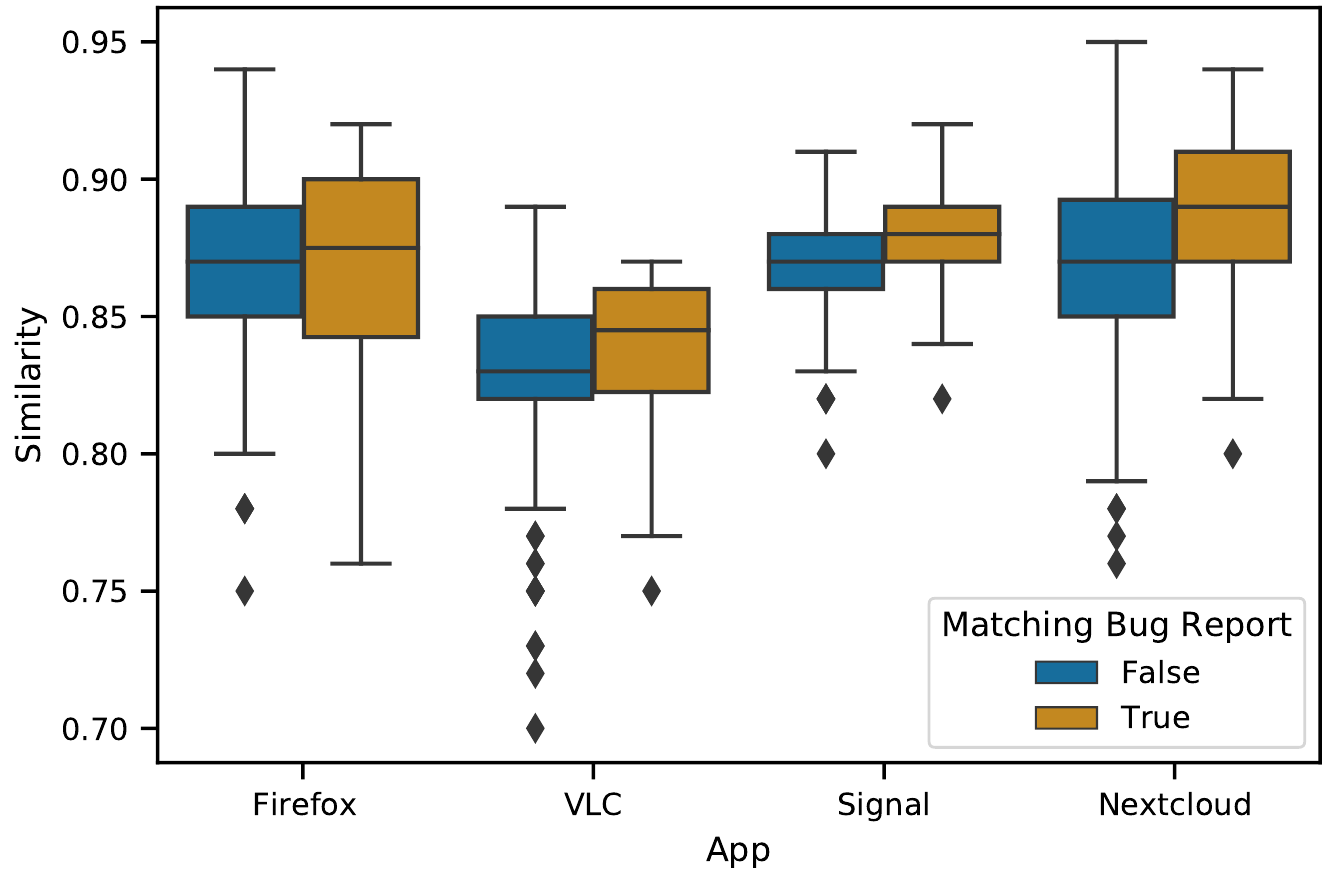}
    \caption{Similarity values for relevant and irrelevant matches per app.}
    \label{fig:similarity}
\end{figure}

\vspace{6pt}
\textbf{Cosine Similarity Analysis}.
We analyzed the cosine similarity values of relevant bug report suggestions and the irrelevant bug report suggestions.
Figure \ref{fig:similarity} shows the cosine similarity values for the manual labeled suggestions for each app.
It shows that the medians of the similarity scores of relevant bug report matches are higher than the irrelevant matches.
However, the similarity scores vary between their min and max values by up to $\sim 0.15$.
All similarity scores are overall high ($\geq 0.5$) as all texts are in the technical domain.

We found that VLC has the lowest cosine similarity score compared to the other apps, which is also the app for which \tool found the fewest relevant bug report matches (26 matches).
The lower cosine similarity indicates a higher language gap between VLC problem reports and bug reports.
To further analyze this indication, we calculated the overlap of nouns used in problem reports with the nouns used in bug report summaries.
We only checked the noun overlap as this is the part-of-speech category \tool uses to generate matches.
For each app, we calculated the ratio between the number of nouns used in problem reports and bug reports, and the number of nouns used overall.
The apps' ratios are: Firefox 19\%, VLC 11\%, Signal 24\%, and Nextcloud 25\%.
The noun overlap calculation strengthens our assumption that the language between the VLC problem reports and the bug report summaries diverge more than the other apps, which negatively affects \tool's automatic matching approach.

\subsection{Qualitative Analysis of \tool's Relevant, Wrong, and Missed Suggestions (RQ2)} \label{ssec:result_qualitative}
We summarize and describe qualitative insights to learn about \tool's relevant and irrelevant suggestions.
Table \ref{tab:examples} provides examples of problem reports, \tool suggested bug report summaries, and our coding of whether we think that there is a relevant match for developers.
In the table, we selected one problem report per app and searched for cases that highlight some of our insights, like recurring bug reports for Signal, or a problem report submitted long before a bug report was filed in the Nextcloud app. 
In the following, we discuss our findings.

\begin{table*}[]
\scriptsize
\centering
\caption{Example problem reports from app reviews and \tool's suggested matching bug reports. The relevant column shows whether the two coders annotated the suggestions as relevant for developers.}
\label{tab:examples}
\begin{tabular}{@{}l|l|c@{}}
\toprule
\multicolumn{1}{c|}{Problem Report}                                                                                                                                                                                                                                                                                                             & \multicolumn{1}{c|}{Suggested Bug Report Summary}                                                                                                                                                    & \multicolumn{1}{l}{Relevant} \\ \midrule
                                                                                                                                                                                                                                                                                                                                                & {\setlength{\extrarowheight}{1.4mm}\begin{tabular}[c]{@{}l@{}}Date: 2018-01-04\\ Report: Only happening with latest version, But keep getting FFbeta draining battery too fast\vspace{1.4mm}\end{tabular}}                               & yes                          \\
                                                                                                                                                                                                                                                                                                                                                & {\setlength{\extrarowheight}{1.4mm}\CC \begin{tabular}[c]{@{}l@{}}Date: 2017-11-20\\ Report: The topbar on android phone becomes white, which makes the time and battery life invisible.\vspace{1.4mm}\end{tabular}} & \CC no   \\
\multirow{-11}{*}{\begin{tabular}[c]{@{}l@{}}App: Firefox\\ Date: 2020-04-21\\ \\ Report:\\ Latest update started consuming over\\ 80\% battery. Had to uninstall to even\\ charge the phone!\vspace{1.4mm}\end{tabular}}                                                                                                                                       & {\setlength{\extrarowheight}{1.4mm}}{\setlength{\extrarowheight}{1.4mm}\begin{tabular}[c]{@{}l@{}}Date: 2016-12-13\\  Report: "Offline version" snackbar is displayed when device is very low on power and in battery saving mode\vspace{1.4mm}\end{tabular}}                 & no                           \\ \midrule

                                                                                                                                                                                                                                                                                                                                                &  \CC {\setlength{\extrarowheight}{1.4mm}\begin{tabular}[c]{@{}l@{}}Date: 2015-09-13\\ After update: no notification sent with TextSecure message. I have to open the app to see if there's something new\vspace{1.4mm}\end{tabular}}                                                     & \CC yes                           \\
                                                                                                                                                                                                                                                                                                                                                & {\setlength{\extrarowheight}{1.4mm}\begin{tabular}[c]{@{}l@{}}Date: 2016-01-17\\ Report: No notifications show up until the app is manually open\vspace{1.4mm}\end{tabular}}                                          & yes  \\
\multirow{-9.5}{*}{\begin{tabular}[c]{@{}l@{}}App: Signal\\ Date: 2017-10-09\\ \\ Report:\\ it is a good app. i am mostly satisfied with\\ it but sometimes, the notifications would not\\ work; so, I would not know that someone \\messaged me until I open the app. it might \\have been fixed because it hasn't been \\happening in the last month or so. Would \\recommended.\end{tabular}}       &  \CC {\setlength{\extrarowheight}{1.4mm}\begin{tabular}[c]{@{}l@{}}Date: 2016-04-17\\  Report: Not getting notification in real time unless I open the app\vspace{1.4mm}\end{tabular}}              &  \CC yes                           \\ \midrule
                                                                                                                                                                                                                                                                                                                                                &  {\setlength{\extrarowheight}{1.4mm}\begin{tabular}[c]{@{}l@{}}Date: 2019-04-13\\ Report: android navigation bar, shown after a click, shifts and resizes full-screen video\vspace{1.4mm}\end{tabular}}              &  no   \\
                                                                                                                                                                                                                                                                                                                                                &  \CC {\setlength{\extrarowheight}{1.4mm}\begin{tabular}[c]{@{}l@{}}Date: 2018-09-27\\ Report: Play/pause button icon is not shifting while pausing the audio on notification area\vspace{1.4mm}\end{tabular}}                                 &  \CC no                           \\
\multirow{-9}{*}{\begin{tabular}[c]{@{}l@{}}App: VLC\\ Date: 2020-05-17\\ \\Report:\\ So many bugs... Plays in background, but\\ no controls in notifications. When you tap\\ the app to bring up the controls, the video\\ is a still screen. Navigating is a pain.\\ Resuming forgets my place constantly.\\ Basically unusable\vspace{1.4mm}\end{tabular}} & {\setlength{\extrarowheight}{1.4mm}\begin{tabular}[c]{@{}l@{}}Date: 2013-09-16\\ Report: {[}Android{]} On video playing the navigation bar is not hidden on some tablets\vspace{1.4mm}\end{tabular}}                 &  no   \\ \midrule
                                                                                                                                                                                                                                                                                                                                                &  \CC {\setlength{\extrarowheight}{1.4mm}\begin{tabular}[c]{@{}l@{}}Date: 2016-07-09\\  Report: nextcloud android client can't login but android webdev clients do\vspace{1.4mm}\end{tabular}}                                                     &  \CC no                           \\
                                                                                                                                                                                                                                                                                                                                                & {\setlength{\extrarowheight}{1.4mm}\begin{tabular}[c]{@{}l@{}}Date: 2018-07-20\\ Report: AutoUpload stuck on "Waiting for Wifi" when using VPN\vspace{1.4mm}\end{tabular}}                                          & yes  \\
\multirow{-9}{*}{\begin{tabular}[c]{@{}l@{}}App: Nextcloud\\ Date: 2017-10-09\\ \\ Report:\\ I have a nextcloud server and the way I\\ access my server is via OpenVPN. The\\ problem now is the  nextcloud native app\\ doesn't work through vpn. It is an odd\\ behavior.  I highly recommend to use\\ owncloud app instead.\end{tabular}}       &  \CC {\setlength{\extrarowheight}{1.4mm}\begin{tabular}[c]{@{}l@{}}Date: 2020-06-18\\  Report: SecurityException in OCFileListAdapter: uid 10410 cannot get user data for accounts of type: nextcloud\vspace{1.4mm}\end{tabular}}              &  \CC no                           \\ \bottomrule
\end{tabular}
\end{table*}

\vspace{6pt}
\textbf{Strength of Contextual Embeddings}.
One strength of our approach is to learn the context of words (which words belong together).
Other approaches like bag-of-words or tf-idf do not consider the context of words and, therefore, fall short in representing a deeper understanding of the language.
The following two examples illustrate the strength of word context.
\tool suggested matches that included the phrases ``automatic synchronization'' and ``auto upload'' in Nextcloud bug reports, as well as ``download pdf'' and ``save pdf'' for the Firefox browser.
One full example is shown in Table \ref{tab:examples}.
The Firefox problem report discusses a ``consuming battery'' problem that happens since the ``latest update''.
The relevant matching bug report states that the ``battery draining'' becomes a problem in the ``latest version''.
It shows that the contextual embeddings of the noun tokens, e.g.,  ``synchronization'' and ``upload'' reach a high text similarity score as they are considered closely related.

\vspace{6pt}
\textbf{Language Gap Leads to Fewer Relevant Matches}.
During the manual coding task, we noticed that the phrasings in VLC's bug reports often contain technical terms, for example: \emph{``Freeze entire android OS when playing a video. libvlc stream: unknown box type cTIM (incompletely loaded)''}.
However, users are typically not part of the development team and do not include technical words like specific library names used by the developers.
Our previously reported plot of the cosine similarity values in Figure \ref{fig:similarity} quantitatively indicated that there might be a language gap as the text similarity scores between problem reports and bug reports were the lowest for VLC.
We then performed a noun overlap analysis, which strengthened the indicator for the language gap as VLC has the lowest noun overlap with 11\%.
Eventually, we looked into the problem reports, bug reports, the Google Play Store, and the issue trackers.

We found that the developers of Nextcloud sometimes reply to problem reports in the Google Play Store and ask the users to also file a bug report in the issue tracker systems.
We do not know how many users are actually going to the issue tracker to report a bug.
But this could also explain why Nextcloud has the highest cosine similarity score and highest noun overlap (25\%).
Consequently, \tool is more accurate if bug report summaries contain non-technical phrases, as users rarely use technical terms.

Sometimes users do not understand the app features.
The following example from a Signal problem report shows a user confusing a feature with a bug: ``Works well but gives me 2 check marks immediately after sending my text. I know the receivers are not reading the texts so fast. Why 2 checkmarks?''
The two checkmarks in Signal are shown as soon as the addressed user successfully received the message.
Signal has an optional feature that changes the color of the two checkmarks to blue if the recipient reads the message.


\vspace{6pt}
\textbf{Recurring Bug Reports}.
Table \ref{tab:examples} shows an example of recurring bug reports.
The problem report of the Signal app states that the user did not receive notifications of incoming messages.
We considered all three matching bug report suggestions of \tool as relevant, as they state the same problem.
The interesting insight in this example is that with \tool we were able to identify a recurring bug report, as the first one was filed in September 2015, the second in January 2016, the third in April 2016.
The problem report of the user happened in October 2017, more than one year after the last suggested match.
In Section \ref{ssec:discuss_recurring}, we discuss how \tool can help systematically find such cases.

\vspace{6pt}
\textbf{Date Case Analysis}.
Regarding the date analysis, we found that in 35 of 167 relevant matches, developers reported the bug reports after the users submitted the corresponding review in the Google App Store.
The time differences of the 35 cases in which the problem report submission happened before the bug report, is 490 days later, on average.
In the following, we illustrate three examples.

Table \ref{tab:examples} shows one problem report for Nextcloud, submitted in October 2017, while the matching bug report was filed in July 2018.
Another user submitted the following problem report on the Nextcloud app: \emph{``Autoupload not working, android 7, otherwise all seems good. Happy with app and will increase stars to 5 when auto upload is working.''}
\tool identified the matching bug report \emph{``Android auto upload doesn't do anything''} that was created 29 days after the problem report.
In the last example, a developer documented a matching bug report 546 days after the corresponding problem report for the Signal app.
Both the user and the developer address the in-app camera feature:
\emph{``Newest update changes camera to add features, but drastically reduces quality of photos. Now it seems like the app just takes a screenshot of the viewfinder, rather than taking a photo and gaining from software post-processing on my phone. [...]''}.
The bug report stated: \emph{``In-app camera shows different images for preview and captured''}.

\section{Discussion} \label{sec:discussion}
This section discusses potential use cases of \tool to support developers in their software evolution process.

\subsection{Detecting Bugs Earlier}
It is essential for app developers to address users' problems as their dissatisfaction may lead to the fall of previously successful apps \cite{Li:Satisfaction:2010, Williams:App:2018}.
One way to cope with user satisfaction is to quickly fix frustrating bugs, which may cause users to switch to a competitor and submit negative reviews.
However, bugs may occur for different reasons.
Some bugs affect only a few users with specific hardware or software versions, while others affect a large user group.
Further, not all bugs are immediately known to developers, particularly, non-crashing bugs, which are hard to discover in automated testing and quality assurance \cite{Martens:RE:2019}.
Our results show that some users submit problem reports in the Google Play Store months before developers document them as bug reports in the project's issue tracker.
When considering additional feedback channels such as social media and other stores, this might get even worse.

Our qualitative analysis of bug reports shows that these earlier submitted problem reports contain valuable information for app developers, such as the affected hardware.
Therefore, we emphasize that developers should continuously monitor user feedback in app stores to discover problems early and document them as bug reports in their issue trackers \cite{martens2019release}.
For this purpose, developers can first apply the automatic problem report classification of app reviews and subsequently use \tool to find existing matching bug reports.
In case \tool does not find matching bug reports, we suggest that developers should consider the problem report as an unknown bug.
However, to avoid the creation of duplicate bugs, we further suggest checking the issue tracker beforehand.
Mezouar et al. \cite{Mezouar:EMSE:2018} suggest a similar recommendation for developers when considering tweets instead of app reviews.
They show that developers can identify bugs 8.4 days earlier for Firefox and 7.6 days earlier for Chrome (on average).

We envision different ways to suggest new bugs to developers. 
First, we could build a system that shows newly discovered bugs to developers. 
From that system, developers can decide to file a new issue in the issue tracker, delay, or reject it.
Alternatively, a bot can, e.g., file a new issue in the issue tracker systems automatically \cite{Martens:RE:2019}.
For the latter, future research could develop, e.g., approaches could prepare certain text artifacts, including steps to reproduce, meaningful issue description, or context information in a template for creating a new issue.

Furthermore, \tool's application is not limited to user feedback in the form of app reviews.
Our approach can generally process user feedback on various software, which developers receive via different channels, including app stores, social media platforms like Twitter and Facebook, or user support sites.
\tool's main prerequisite is written text.

\subsection{Enhancing Bug Reports with User Feedback}
Martens and Maalej \cite{Martens:RE:2019} analyzed Twitter conversations between vendors' support accounts like @SpotifyCares and their users. 
Similarly to our statement, the authors highlight that users who provide feedback via social media are mostly non-technical users and rarely provide technical details. 
As support teams are interested in helping users, they initiate a conversation to ask for more context and details. 
They ask for context information like the affected hardware device, the app version, and its platform. 
Their objective is to better understand the issue to potentially forward that feedback to the development team and provide more helpful answers.
Hassan et al. \cite{hassan2018studying} show that developers also communicate with their users in the app stores to better understand their users.

Zimmermann et al. \cite{Zimmermann:TSE:2010} show that the most important information in bug reports are steps to reproduce, stack traces, and test cases. 
The participants of their survey found that the version and operating system have lower importance than the previously mentioned information. 
However, the authors also argue that these details are helpful and might be needed to understand, reproduce, or triage bugs.
Nevertheless, the authors did not focus on apps but developers and users of Apache, Eclipse, and Mozilla.

Developers could further use \tool to understand the popularity of bugs. 
They can achieve this in two steps.
First, change \tool to take bug reports as an input to suggest problem reports (inverting the order as reported in the approach).
Second, the parameter for the number of suggestions can either be increased or removed to enable suggesting all problem reports sorted by the similarity to the given bug report.
This leads to an aggregated crowd-based severity level, a bug popularity score, or an indicator of how many users are affected by a certain bug report.

We further envision extracting context information and steps to reproduce from user feedback to enhance the issue tracker's bug report description.
Having this information at hand can help developers narrow down the location of an issue and understand how many users are affected. 
Developers can use \tool to find problem reports related to bug reports by simply using a bug report summary as the input in our approach.
Then, developers can skim through the suggested problem reports, select those that seem relevant, and then check whether they contain relevant context information.
In case users did not provide useful information, developers can take the IDs of the relevant problem reports and request more information from users in the Google Play Store.
This process can partly be automated, e.g., using bots. 

\subsection{Extending \tool to Identify Duplicated, Recurring, or Similar Bug Reports} \label{ssec:discuss_recurring}
In Section \ref{sec:results}, we found that \tool identified recurring bug reports.
The Signal example in Table \ref{tab:examples} shows a recurring bug report.
Within the three bug report suggestions, \tool found three relevant matches.
While the first bug report was filed in September 2015, the second in January 2016, the third in April 2016, a user reported the problem again in October 2017.

Consequently, developers might want to adapt \tool to either find recurring, similar, or duplicated bug reports even though it is not \tool's primary goal.
However, since the approach evaluates the matches based on context-sensitive text similarity, it could lead to promising results.
Developers interested in these cases could, for example, increase \tool's parameter \emph{number of matching bug report suggestions} and use a bug report summary as the input for \tool to identify these cases.
Future work could investigate and evaluate the use of \tool for such cases by utilizing our replication package.
\section{Threats to Validity} \label{sec:validity}
We discuss threats to internal and external validity.
Concerning the internal validity, we evaluated \tool by manually annotating 600 suggested bug reports for 200 problem reports.
We performed two annotation tasks. 
One task to verify that the automatically classified app reviews are problem reports, and one to annotate whether \tool's suggested matches are relevant for developers.
As in every other manual labeling study, human coders are prone to errors.
Additionally, their understanding of ``a relevant match'' may differ, which could lead to disagreements.
To mitigate this risk, we designed both annotation tasks as peer-coding tasks.
Two coders, each with several years of app development experience, independently annotated the bug report matches.
For the verification of problem reports, we used a well-established coding guide by Maalej et al. \cite{maalej2016automatic}, which Stanik et al. \cite{Stanik:Feedback:2019} also reused for the automatic problem report classification. 
To mitigate the threat to validity regarding the annotation of relevant matches, we performed test iterations on smaller samples of our collected dataset and discussed different interpretations and examples to create a shared understanding.

Further, we tried to collect a representative sample of meaningful app reviews.
Thereby, we collected up to 10,000 app reviews for each app, ordered by helpfulness, covering more than two years.
We did not aim for a comprehensive app review sample for a specific time frame but prioritized a meaningful app review sample from a larger time frame.
Thereby, we could identify diverse insights within our qualitative analysis.

Another potential limitation is that we only considered 50 app reviews per app (200 in total), which we automatically classified as problem reports.
This classification might only find a specific problem report type, neglecting other informative problem reports.
Other kinds of app reviews, including feature requests or praises, might also contain valuable information for developers, which \tool could match to bug reports.
Therefore, our observations might differ for another sample of app reviews.

In the case \tool could not find any matching bug report among the three suggestions, we manually searched for relevant bugs in the issue trackers.
We queried different term combinations and synonyms for certain features and components similar to how developers would proceed.
However, not finding a relevant match in the issue tracker systems does not prove the non-existence of a relevant bug report in the issue tracker as we could have missed important terms in the query.

Concerning the external validity, our results are only valid for the four open source apps of our dataset.
We considered different app categories, covering many tasks that users perform daily by including Firefox as an app for browsing the internet, Signal for messaging, Nextcloud for cloud storage, and VLC as a media player for music, videos, and streaming.
However, these app categories include popular apps that we do not cover in our study, like Chrome or Safari. 
Further, the bug report suggestions could differ for closed source projects or apps of other mobile operating systems.
\section{Related Work} \label{sec:related_work}
\subsection{User Feedback Analytics}
Feedback-driven requirements engineering is an increasingly popular topic in research often focusing on \emph{app reviews} \cite{harman2012app, guzman2014users, maalej2016automatic}, \emph{tweets} \cite{guzman2016needle, Williams:Twitter:2017}, product reviews such as  \emph{Amazon reviews} \cite{kurtanovic2017mining, kurtanovic2018user}, or a combination of reviews and product descriptions \cite{johann2017safe}.
All of them have in common that a software product already exists and that users rate and write their experience with it after using it \cite{Pagano:App:2013}.
User feedback and involvement are important to both software engineers and requirements managers, as they often contain insights such as introduced bugs and feature requests \cite{Villarroel:Release:2016, maalej2016toward, Stanik:Feedback:2019}.
The classification of user feedback \cite{maalej2016automatic} was a first step towards understanding user needs.
Further studies \cite{kurtanovic2017mining, kurtanovic2018user} looked at the classified feedback more closely by analyzing and understanding user rationale---the reasoning and justification of user decisions, opinions, and beliefs.
Once a company decides to integrate, for example, an innovative feature request in the software product, it will be forwarded to the release planning phase \cite{nayebi2018escalation, nayebi2018asymmetric}.

In our approach, we build on top of the existing body of research by, in particular, applying the machine learning approach of Stanik et al. \cite{Stanik:Feedback:2019} to identify problem reports in app reviews.
We used that approach as an initial filter of the app reviews because Pagano and Maalej \cite{Pagano:App:2013} showed that popular apps receive about 4,000 app reviews daily---which would be unfeasible for us to filter manually.
Since the classification approach has an F1-score of 0.79 for identifying problem reports in English app reviews, we had to manually check the classified app reviews, as described in Section \ref{sec:study_data}.

\subsection{Combining User Feedback and Bug Reports}
El Mezouar et al. \cite{Mezouar:EMSE:2018} present a semi-automatic approach that matches tweets with bug reports in issue tracking systems.
They look at the bug reports of the two browsers Firefox and Chrome.
They use natural language processing techniques to preprocess the text of both data sources and apply the Lucene search engine to suggest potentially matching bug reports.
The approach crawls, preprocesses, filters, and normalizes tweets before they match them with issues.
During the crawling process, the authors include tweets that either mention the browser with the @ symbol or hashtag.
Then, they remove misspellings, abbreviations, and non-standard orthography.
Afterward, the authors filter tweets with a list of bug related keywords like \emph{lag} and \emph{crash} while also considering negated bug terms with a part-of-speech analysis.
In a final step, the approach removes symbols, punctuation, non-English terms, and stems the words using the porter stemmer \cite{porter1980algorithm}.
For matching tweets with issues, the authors extract keywords from the tweets and use them as a search query in the Lucene search engine\footnote{\url{https://lucene.apache.org}}.

In contrast to El Mezouar et al., we consider app reviews from App Stores.
While Tweets allow for lengthy conversations with stakeholders \cite{guzman2017little, Martens:RE:2019} that may lead to in-depth insights into, e.g., the users' context like the app version and steps to reproduce, app stores enable developers to reply to app reviews, and users to update their review \cite{hassan2018studying}.
App reviews also contain metadata like the hardware device and the installed app version (that information is only available to the app developers).
Further, stakeholders can ensure that users address the app the user wrote reviews for.
When analyzing tweets and the software is available on multiple platforms like Windows, Mac, iOS, and Android, it is often difficult to understand which platform the user addressed without interacting with the user.
Third, in app reviews, users can write longer texts than in tweets.
Besides considering two platforms as our data source, we further applied more sophisticated technical solutions by applying state of the art NLP approaches that have a deeper understanding of the language than a search engine.
We also build on top of previous research to extract problem reports from app reviews, leading to more relevant results than a simple keyword-based approach \cite{maalej2016automatic}.
\section{Conclusion} \label{sec:conclusion}
In this paper, we introduced \tool, an approach that extracts problem reports from app reviews submitted by users; and then identifies matching bug reports in an issue tracker used by the development team.
Our approach primarily addresses the challenge of integrating user feedback into the bug fixing process.
Developers may receive thousands of app reviews daily, which makes a manual analysis hard to unfeasible.
Additionally, most user feedback is either praise like ``I love this app.'' or a dispraise like ``I hate it!''.
For the latter reason, we first filtered the problem reports from the reviews by reusing recent related work.
After manually validating the problem reports, we applied \tool, which takes a problem report and a bug report summary as the input.
\tool then transforms the text into context-sensitive embeddings on which we applied cosine similarity to identify potential matching problem reports and bug reports.
In total, from 200 problem reports, \tool was able to identify 167 relevant matches with bug reports.
In 91 cases, \tool did not find any match.
To understand whether indeed no match exists, we manually looked into corresponding issue trackers and found that in 44 cases, \tool missed a potential match while in 47 cases, no bug report existed.
Our results show that our approach can help developers identify bugs earlier, enhance bug reports with user feedback, and eventually lead to more precise ways to detect duplicate or similar bugs. 
\section*{Acknowledgment} 
This work was partly supported by the City of Hamburg (within the ``Forum 4.0'' project as part of the ahoi.digital funding line), by the European Union (within the Horizon 2020 EU project ``OpenReq'' Grant Nr. 732463), and by the German Federal Government (BMBF project ``VentCore'').

\bibliographystyle{abbrv}
\bibliography{references/bib}
\end{document}